\documentclass[aps,,english,10pt,superscriptaddress,twocolumn,
tightenlines, notitlepage]{revtex4-1}
\usepackage[section]{placeins}
\usepackage{relsize}
\usepackage{amsfonts}
\usepackage{hyperref}
\usepackage{bbm}
\usepackage{mathtools}
\usepackage{tikz}
\usepackage{physics}
\usepackage{amsthm}
\usepackage{amsmath}
\usepackage{amssymb}
\usepackage{tensor}
\usepackage{bigints}
\usepackage{mdframed}
\usepackage{xcolor}
\usepackage{graphicx}
\usepackage{float}
\usepackage{booktabs}

\DeclareGraphicsExtensions{.png,.tikz}
\graphicspath{{./Data/img}}

\newcommand{\qexp}[1]{\left\langle#1\right\rangle}

\newcommand{\qvar}[1]{\langle(\Delta#1 )^2\rangle}
\newcommand{\spin}[1]{\hat{S}^{#1}}

\begin{document}

\title{Revealing divergent length scales using 
quantum Fisher information in the Kitaev honeycomb model}
\date{\today}
\author{James Lambert}
\email{lambej3@mcmaster.ca}
\author{Erik S. S{\o}rensen}
\email{sorensen@mcmaster.ca}
\affiliation{Department of Physics \& Astronomy, McMaster University
1280 Main St.\ W., Hamilton ON L8S 4M1, Canada}
\begin{abstract}
  We compute the quantum Fisher information (QFI) associated with two different local
  operators in the ground state of the Kitaev honeycomb model, 
  and find divergent behaviour in the
  second derivatives of these quantities with respect to the driving parameter at
  the quantum phase transition between the gapped and gapless phases for both
  fully anti-ferromagnetic and fully ferromagnetic exchange couplings, thus
  demonstrating that the second derivative a locally defined, experimentally
  accessible, QFI can detect
  topological quantum phase transitions.
  The QFI associated with a local magnetization operator behaves differently
  from that associated with a local bond operator
  depending on whether the critical point is approached from the gapped or
  gapless side. We show how the behaviour of the second derivative of the QFI at
  the critical point can be understood in terms of the diverging length scales
  associated to the two and four point correlators of
  the Majorana degrees of freedom. We present critical exponents associated with the 
  divergences of these length scales. 
\end{abstract}

\maketitle

\section{Introduction}\label{sec:Intro}
\subsection{Overview}\label{subsec:Intro:Overview}

The quantum Fisher information (QFI), $\mathcal{F}$, arises naturally in quantum metrology
~\cite{Petz1996,Petz2002,Paris2009,Toth2012,Toth2013,Toth2014}. Given a general
state $\rho(\theta)$ where $\theta$ is some parameter, the QFI 
bounds the precision with which $\theta$ may be extracted in \emph{any}
$M$ measurements through the
Cram\'{e}r-Rao bound, $\text{Var}_{\hat{\rho}}(\theta_i) \geq 1
/\sqrt{M\mathcal{F}}$. 
In other words the QFI quantifies the extent to which a parameterized state $\rho(\theta)$ may
be distinguished from a neighbouring state $\rho(\theta+\dd\theta)$. By
quantifying the distinguishability of neighbouring states, 
the QFI furnishes a natural notion of distance on the Hilbert space, with more
easily distinguishable states separated by a greater distance. Formally, the QFI
quantifies the local change in the Bures distance under the aforementioned
parameterization.~\cite{holevo2003statistical,bengtsson2017geometry,wootters1981statistical,braunstein1994statistical}
This geometrical interpretation of the QFI expands its scope of
application to probing the physics of condensed matter phases and phase
transitions.~\cite{InformationTheoreticApproach,NaturePhysics,Liu2013,Zheng2015,Ma2009,lambert2019estimates,
yang2008fidelity} The QFI also exhibits interesting behaviour during a quantum
quench in spin chain systems~\cite{PhysRevA.101.062105} In fact, a special case of the QFI is already ubiquitous in theoretical studies
of condensed matter systems. The fidelity
susceptibility~\cite{zanardi2006ground} (FS) is directly
proportional to the QFI~\cite{liu2014fidelity} for an appropriate
parameterization. In particular 
One often considers parametrizations that have been
generated unitarily (though this is not the only choice) 
by a Hermitian operator $\hat{O}$ according to,
$\rho(\theta) = e^{i\theta\hat{O}}\rho e^{-i\theta\hat{O}}$, and we will
restrict ourselves to this case here. The operator $\hat{O}$ is usually
expressed as a sum over sub-operators $\hat{O}^\alpha$,
\begin{equation}
  \hat{O} = \sum_r \hat{O}^{\alpha_r}_r
  \label{}
\end{equation}
A \emph{local} operator, is one for which all $\hat{O}_r^{\alpha_r}$ depend on a
contiguous sublattice that is small relative to the total lattice. One may also
consider \emph{non-local} parameterizations, such as the string operators
considered in Ref.~\cite{pezze2017multipartite}. Non-local
parameterizations reveal remarkable behaviour in topological phases as
demonstrated in Ref.~\cite{pezze2017multipartite}, where a characteristic, 
super-extensive scaling
of the non-local QFI is demonstrated in the topologically non-trivial phases of the
Kitaev wire.  

In a many-body state containing $N$ degrees of freedom, the QFI density
$F=\mathcal{F}/N$ quantifies the degree of multipartite entanglement when the
state $\rho$ is projected into the eigenbasis of the operator $\hat{O}$ that
generates the parametrization. For $F>m$ where $m|N$, we say the state is
$(m+1)$ partite entangled.~\cite{Toth2012,hyllus2012fisher, Pezze2009}
Specifically for the case of pure states, $\psi$, and unitary parametrizations, the QFI is
proportional to the variance of the generator,~\cite{holevo2003statistical}
\begin{equation}
  \mathcal{F} = 4\text{Var}_\psi(\hat{O})
  \label{eq:QfiPureUnitDefn}
\end{equation}
While one could perform an interesting study looking only at the variances, we
prefer to work within the context of the QFI because it continues to be well
defined at finite temperature. This link allows the critical properties of the
ground state to be inferred from the thermal scaling of the QFI.~\cite{NaturePhysics, gabbrielli2018} 
While we do not consider finite temperature behaviour in this study, the
connection offers a path forward for future work.
Recently it was shown that the QFI can be detected experimentally in inelastic
scattering measurements.~\cite{NaturePhysics} Thus working within the context of
the a locally defined QFI also allows for connection with experiment. This
contributes to a growing body of research on experimental approaches to extract
multipartite
entanglement.~\cite{PhysRevLett.106.020401,PhysRevLett.103.100502,PhysRevB.89.125117}
We emphasize that we do not detect genuine multipartite entanglement in this
study, but mention the connection for completeness. 

Given that the QFI is defined at finite temperature and that the zero temperature
QFI is proportional to the variance, one might ask, what is the generalization of the notion of
variance to the finite temperature case? By imagining that the variance
of an observable contains a
quantum contribution and a thermal contribution,  
quantum variances (QV) may be defined which are proportional to an upper and
lower
bound of the QFI.~\cite{frerot2016quantum} The QFI at zero temperature can be
viewed as the zero temperature limit of the quantum contribution to the
variance (the thermal contribution being zero at zero temperature).

The QFI has now been studied in a wide range of
models.~\cite{NaturePhysics,Liu2013,Zheng2015,Ma2009,lambert2019estimates} Of
particular interest for our purposes is the work done on the Kitaev wire
in Ref.~\cite{pezze2017multipartite}, where the first derivative of the QFI associated with a
local generator was shown to exhibit a divergence at the topological phase
transition of that model, and where the topologically non-trivial phase
exhibits super-extensive scaling of the QFI associated with a non-local
generator.  

Quantum spin liquids (QSL) are characterized by a lack of any form of long range
magnetic order down to zero temperature.~\cite{QSLReview} Such phases are
thought to exhibit instead subtle forms of quantum ordering, along with
topologically non-trivial anyonic excitations.~\cite{wen2002quantum} In so far
as these phases are characterized by a lack of order, their detection in
experiment presents a substantial challenge. In this work we examine the
behaviour of the QFI in the Kitaev honeycomb model
(KHM)~\cite{AnyonsExactlySolved}, which presents two spin liquid phases (one
gapped and one gapless), induced by exchange coupling anisotropy. 

The KHM has been studied from an information theoretic perspective before, with
studies examining the Jensen-Shannon
divergence~\cite{chen2019topological}, and the mutual
information.~\cite{cui2010quantum} Of particular interest is the Fidelity
susceptibility, which was studied in Ref.~\cite{yang2008fidelity} and the study of the
Bures distance in.~\cite{abasto2009thermal} In the case of an $n$ parameter
estimation scenario (or an $n$ dimensional unitary parametrization), the Bures
distance is locally equivalent to the QFI Matrix which is a Riemannian metric on
the Hilbert space.~\cite{holevo2003statistical} The Fidelity susceptibility is recovered by examining the
particular parametrization of the Hilbert space corresponding to the driving
operator of the phase transition. The physics of the KHM have also been studied
using SU(2) parton approaches~\cite{PhysRevB.84.125125}. Details of the
dynamical response of the model in the presence of magnetic fields may be found
in Ref.~\cite{PhysRevB.100.144445}

For the remainder of this section we introduce the KHM and explain its key
features. In Sec.~(\ref{sec:QfiSuscepAndDivergingLengthScales}) we discuss the 
relationship between the scaling of the second
derivative of the QFI density (hereafter called the QFI \emph{susceptibility}), $\partial_u^2F$, 
where $u$ drives the phase transition,
and the correlation functions of the generator.
In Sec.~(\ref{sec:QfiSucepForMagAndBond} we analyze the behaviour of these quantities
for the \emph{magnetization} operator, $\sum_j S_j^\alpha$, (where $j$ represents
both a unit cell position and sublattice index), in
Sec.~(\ref{sec:QfiSuscep:MagOp}), and the $\emph{bond}$, $\sum_r S_{r,A}^\alpha
S_{r,B}^\alpha$, where $r$ indicates a unit cell, in
Sec.~(\ref{sec:QfiSuscep:BondOp}). Finally, we conclude our discussion in
Sec.~(\ref{sec:Conclusion}), where we discuss the relevance of this work to
studies of  the geometric phase.

\subsection{Kitaev Honeycomb Model}\label{sec:IntroKHM}
  
  The Kitaev honeycomb model (KHM) is given by,
  \begin{equation}
    H = \sum_{\langle j,k\rangle} K^{\gamma_{j,k}}
    S^{\gamma_{j,k}}_jS^{\gamma_{j,k}}_k
    \label{eq:KitaevModel}
  \end{equation}
  where the sum is over nearest neighbour bonds and $\gamma\in\{x,y,z\}$ denotes a
  bond-dependent Ising exchange. 
  If the exchange couplings are sufficiently
  isotropic ($\qty|K^\gamma| \leq \qty|K^\alpha| + \qty|K^\beta|$, for all
  choices of 
  $\alpha,\beta,\gamma\in\{x,y,z\}$), the spectrum is gapless. In the regime
  where one exchange coupling is dominant (the opposite inequality), the model
  is gapped. This phase transition between two topologically different spin
  liquid phases presents no local order parameter. It is instead associated with
  a subtle kind of symmetry breaking to do with the structure of the gauge
  fields themselves.~\cite{AnyonsExactlySolved} 
  On the gapped side of the transition, the model is mapped
  onto the lattice gauge Ising model~\cite{kardar2007statistical}, 
  with alternative rows of hexagon plaquettes
  becoming associated with one of the two excitations in that model
  (conventionally called $e$ and $m$  excitations). In both phases, the spin-spin 
  correlation functions are identically zero beyond nearest neighbour. 
  The model also possesses an extensive number of conserved charges defined by
  the plaquette operators.

  Remarkably, the KHM is analytically solvable.~\cite{AnyonsExactlySolved} 
  By mapping each spin operator
  into the space of four Majorana fermions, $\{c,b^x,b^y,b^z\}$ via
  \begin{equation}
    S^\gamma_j = \frac{1}{4}ic_jb^\gamma_j\nonumber
    \label{eq:SpinsToMaj}
  \end{equation}
  an extensive number of conserved charges can be constructed, given by $u_{j,k} =
  ib_j^{\gamma_{j,k}} b_k^{\gamma_{j,k}}$. These operators take eigenvalues $\pm1$.   
  Using the above mapping, the KHM becomes,
  \begin{equation}
    H = \frac{i}{4}\sum_{jk} K^{\gamma_{j,k}}u_{jk} c_{j}c_{k}
    \label{}
  \end{equation}
  Since the $u_{j,k}$ commute with the Hamiltonian, we may fix a particular
  configuration of eigenvalues on each bond, and the problem is reduced to free
  Majoranas hopping in the gauge fields. 
  The lowest energy configuration will be the flux free configuration, as
  follows from
  Lieb's theorem.~\cite{lieb2004flux} We therefore choose to work in the configuration where all
  $u_{i,j}$ have eigenvalue $+1$ (hereafter referred to as the standard gauge).
  Once we fix a gauge configuration, the model is a simple hopping Hamiltonian,
  which may be diagonalized by Fourier transforming and then performing a
  Bogoliubov rotation, where the mixing angle is defined implicitly via,
  \begin{equation}
    \tan(2\theta_q) = \frac{\epsilon_q}{\Delta_q},
    \label{eq:BogoAngleDefn}
  \end{equation}
  where, 
  \begin{subequations}
  \begin{align}
    \epsilon_q &= K_\alpha\cos(q_x) + K_\beta\cos(q_y) + K_\gamma \\
    \Delta_q &= K_\alpha\sin(q_y) + K_\beta\sin(q_y)
  \end{align}\label{eqs:AngleComps}
  \end{subequations} 
  Where $\alpha,\beta,\gamma\in\{x,y,z\}$ depending on the choice of which bond
  acts as the unit cell. Here $q_x=\mathbf{a}\cdot\mathbf{q}$ and
  $q_y=\mathbf{a}\cdot\mathbf{q}$ where $\mathbf{a}_1$ and $\mathbf{a}_2$ are
  any choice of translation vectors on the principle lattice and
  $q=\frac{n}{L_x}\mathbf{b}_1 + \frac{m}{L_y}\mathbf{b}_2$, with $L_x$ and
  $L_y$ the side length of the lattice, is a general vector in the reciprocal
  space. 

  We note that the true, physical ground state, must be the
  symmetrized product over all physically equivalent choices of the gauge
  fields (i.e. all choice of the gauge fields resulting in zero flux).
  Following arguments described in Ref.~\cite{baskaran2007exact}, 
  the operators we consider are not dependent on projection into the physical
  subspace at large system sizes.
  Details of the solution to the Kitaev model are provided
  in~\ref{app:CalcVariance:SolnKitaev}.

\section{QFI Susceptibility and Diverging Length
Scales}\label{sec:QfiSuscepAndDivergingLengthScales}

In order to interpret the divergences at the critical point, 
consider a generator, $\hat{O}=\sum_r\hat{O}_r$  where the generators are 
given by a sum over local products of spins,
$\hat{\mathcal{S}}_r = \prod_{j\in\Lambda_r}S_{r+\ell_j}^{\alpha_j}$, where $\Lambda_r$ is
some local, contiguous sublattice. The associated QFI density in a pure state is given by, 
\begin{align}
  F\{\hat{O}\} &=
  \frac{1}{N}\sum_{r_1,r_2}\qexp{\mathcal{S}_{r_1}\mathcal{S}_{r_2}}
  -
  \qexp{\mathcal{S}_{r_1}}
  \qexp{\mathcal{S}_{r_2}}
  \label{eq:GenQFIKitaevSpin}
\end{align}
Through Kitaev's mapping, we may decompose our spin blocks into a component
operating on the flux sector, $\mathcal{B}_r=\prod_{j\in\Lambda_r}
b_{r+\ell_j}^{\alpha_j}$, 
and a component operating on the matter sector,
$\mathcal{C}_r=\prod_{j\in\Lambda_r}c_{r+\ell_j}$,
\begin{equation}
  F\{\hat{O}\} =
  \frac{1}{N}\sum_{r_1,r_2}
  \qexp{\mathcal{B}_{r_1}\mathcal{B}_{r_2}}
  \qexp{\mathcal{C}_{r_1}\mathcal{C}_{r_2}}
  - 
  \qexp{\mathcal{B}_{r_1}}\qexp{\mathcal{B}_{r_2}}
  \qexp{\mathcal{C}_{r_1}}\qexp{\mathcal{C}_{r_2}}
  \label{}
\end{equation}
Now there are three possible values for the flux sector expectation values. If
$\mathcal{B}_r$ is diagonal in the gauge sector, then the contribution from the
gauge sector factorizes and gives an overall prefactor of $\pm 1$. If
$\mathcal{B}_r$ is strictly off-diagonal, but
$\mathcal{B}_{r_1}\mathcal{B}_{r_2}$ has diagonal entries, then the situation is
the same. Finally, it may be the case that $\mathcal{B}_{r_1}\mathcal{B}_{r_2}$
has non-zero diagonal elements only for certain separations. Regardless of which
scenario is realized, the contribution to the QFI from the flux sector will be
independent of $u$, since the gauge fields commute at all points in the phase
diagram. The QFI is then given by a sum over the correlation functions  in the
matter sector Majorana fermions with some prefactor (which might be $\pm1$ or
$0$ as a function of the separation), determined by the situation
above. We adopt the standard ansatz for the matter sector correlations 
\begin{equation}
  \qexp{\mathcal{C}_{r_1}\mathcal{C}_{r_2}} = \Xi(r,u)r^{-a}e^{-\frac{r}{\xi(u)}}
  \label{eq:CorrAnsatz}
\end{equation}
where $a$ is determined by the phase and does not depend explicitly on $u$, and
$\xi(u)$ is a length scale associated with correlations between the Majorana
operators which depends on the position in the phase diagram.

Taking the assumption that, near the critical point,
\begin{equation}
  \xi(u)\sim \qty|u-u_c|^{-\nu}
  \label{}
\end{equation}
one can show that the second derivative of the QFI density must diverge at the
critical point like, 
\begin{equation}
  \partial_u^2F\{\hat{O}\}\sim \qty|u-u_c|^{\nu-2}
  \label{eq:QfiSuscepDiverg}
\end{equation}
Therefor $\Delta_{\hat{O}} = \nu-2$. The QFI and, by extension, the QFI
susceptibility, are, in principle, experimentally accessible probes. In
particular for the case of local generator such as the total magnetization
operator. The QFI can there be used to extract experimentally the scaling of the
correlation length associated with the matter sector of the KHM. This analysis
is similar to the analysis performed in the supplementary materials of
~\cite{NaturePhysics}, where the authors examined the effects of coarse-graining
transformation on the QFI to arrive a scaling hypothesis for the near field and
finite temperature regimes. 

In practice,
experimentally relevant models will not be amenable to the above treatment, as
the KHM acquires additional terms in real materials that break the
integrability of Kitaev's  original solution (e.g. Heisenberg terms and
symmetric off-diagonal terms).~\cite{rau2014generic} In these cases the analysis may
instead be applied to the correlation length of the spin degrees of freedom
directly, and divergences in the QFI susceptibility may still be
linked to the critical exponent for the divergence of a correlation length.

\section{QFI Susceptibility for Magnetization and Bond
Operators}\label{sec:QfiSucepForMagAndBond}

Motivated by the results of ~\cite{pezze2017multipartite}, we compute the QFI
associated with two local operators and examine the second derivatives of those
operators with respect to the driving parameter of the phases transition. In
both cases the second derivative of the QFI is found to diverge. We term this
the QFI \emph{susceptibility}.

Throughout this section we consider a path through the space of exchange
couplings parameterized by $u$, 
\begin{subequations}
  \begin{align}
    K^z &= 1 + u, \\
    K^x &= 1, \\
    K^y &= 1
    \label{subeqs:PathDefns}
  \end{align}
\end{subequations}
For this parametrization, $u_c=1$ represents the critical point between the
gapless phase ($u<1$)  and the gapped phase ($u>1$). The ground state of the
KHM is a function of $u$ and is hereafter denoted as $\psi_0(u)$.
We consider the case of fully ferromagnetic and fully anti-ferromagnetic
exchange couplings for both parametrizations (in which case $K^\gamma\rightarrow
-K^\gamma$). Unless otherwise noted, calculations are carried out for $L_x=L_y=10^4$ with periodic
boundary conditions in a rhombic geometry.

\begin{figure}[t]
  \centering
  \includegraphics[width=0.5\textwidth]{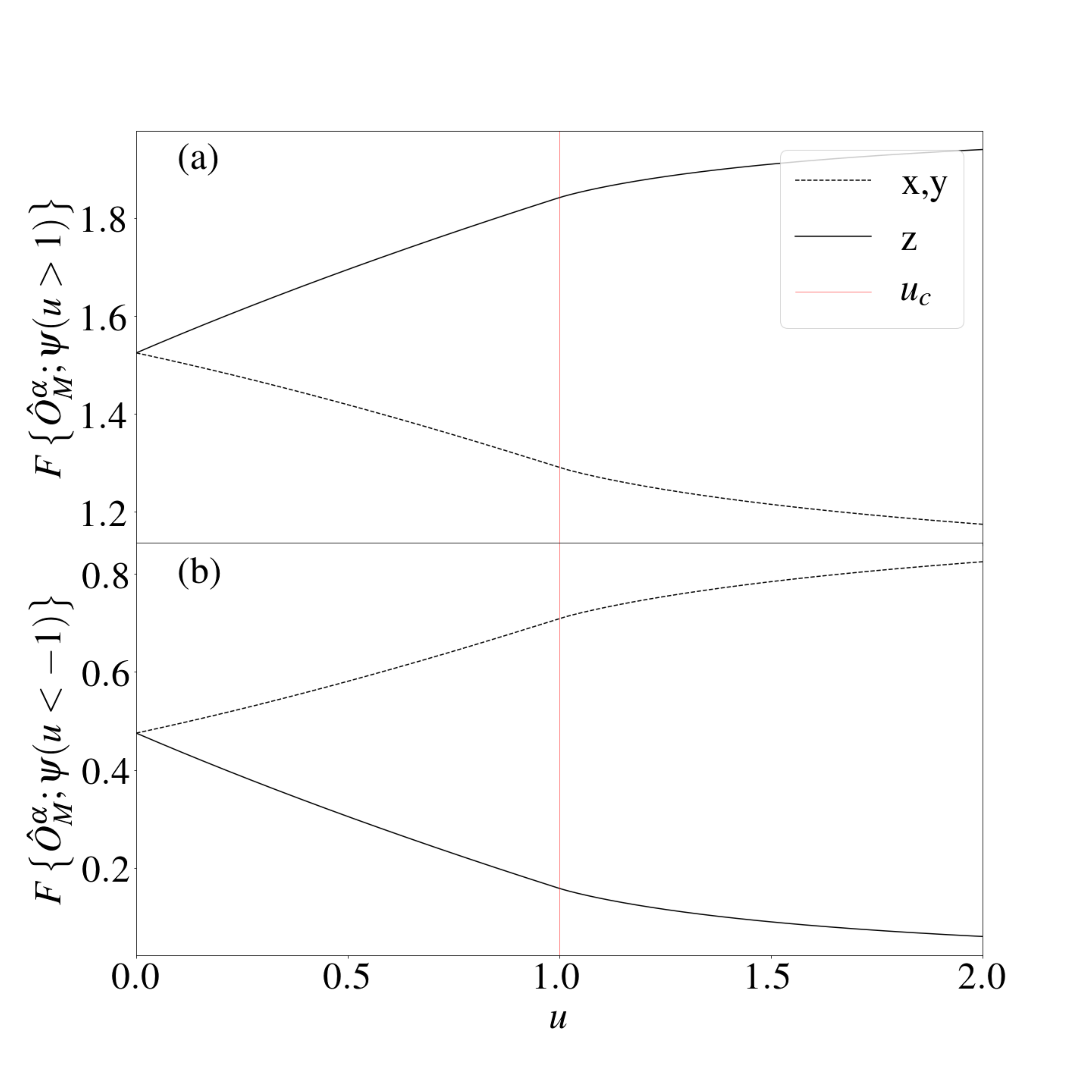}
  \caption{QFI for Magnetization operator with fully anti-ferro (a) and ferro
    (b) magnetic exchange
  couplings. The red vertical line marks the critical value of $u$. Results are
  for $L_x=L_y=10^4$ with a $u$ spacing of $\sim 10^{-3}$.}
  \label{fig:AFMFMfiMagOp}
\end{figure}
\begin{figure}[t]
  \centering
  \includegraphics[width=0.5\textwidth]{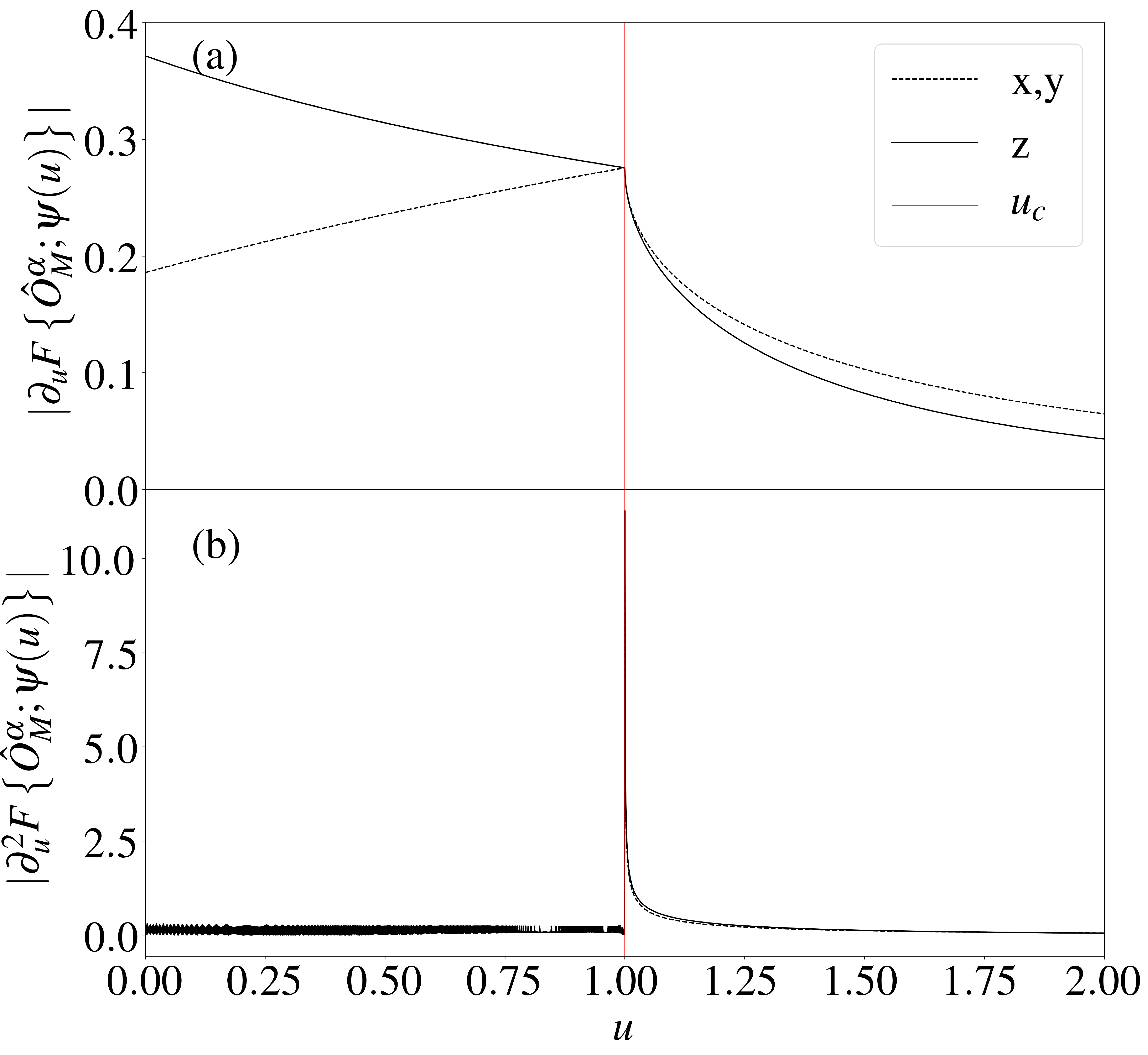}
  \caption{First (a) and second (b) derivatives of the QFI generated by
  the magnetization operator. The results are the same in both the ferro and
anti-ferro magnetic cases. Results for $L_x=L_y=10^4$ with a $u$ spacing
$\sim10^{-3}$. The red line denotes the position of the critical point}
\end{figure}
  \label{fig:MagOpDouble}
  
\subsection{Magnetization Operator}\label{sec:QfiSuscep:MagOp}

First we examine the QFI in the Kitaev honeycomb as generated by the
magnetization operator,
\begin{equation}
  \hat{O}_M^\alpha = \sum_r\left(S_{r,A}^\alpha + S_{r,B}^\alpha\right).
  \label{eq:MagOperDefn}
\end{equation}
here $\alpha\in\left\{x,y,z\right\}$, $r$ denotes a unit cell in the two site
basis, and $A,B$ denotes the sublattice. The corresponding QFI is given by, 
\begin{align}
  F_{M,\alpha}(u) \equiv F\left\{\hat{O}_M^\alpha; \psi_0(u)\right\} 
  &= 4\text{Var}_{\psi_0}(O_M^\alpha) \\ 
  &=1 - \frac{1}{N}\sum_q\cos(2\theta_q) 
  \label{}
\end{align}
with $\psi_0(u)$ defined at the start of this section.

Fig.~\ref{fig:AFMFMfiMagOp} shows this quantity plotted along 
the path defined by~\ref{subeqs:PathDefns} for the fully antiferromagnetic (a)
and ferromagnetic (b) cases
respectively. 
In the AFM case the ground state possesses $F_{M,\alpha} > 1$ for each spin
component. This is indicative of at least bipartite entanglement. In a pure
state any non-zero QFI is indicative of the presence of quantum correlations.
Nonetheless, the QFI in the fully ferromagnetic case is insufficient to witness
even bipartite entanglement, indicating that quantum correlations are reduced
for the FM coupling. 
\begin{figure}[t]
  \centering
  \includegraphics[width=0.5\textwidth]{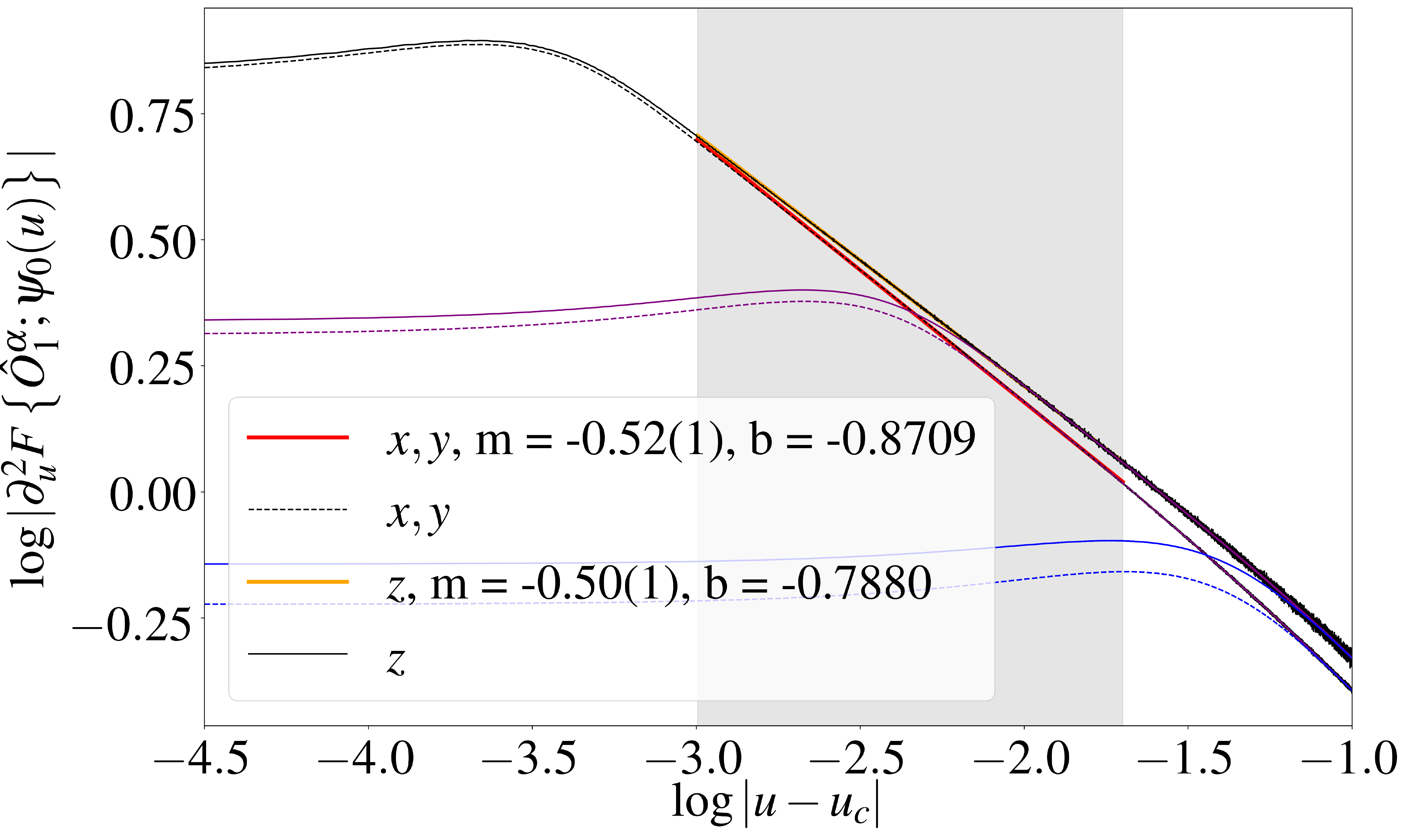}
  \caption{log-log plot of the second derivative of the QFI with respect to the
  magnetization operator and the distance to the critical point from the gapped
  side ($u>1$).
  Results for $L_x=L_y=10^4$ with a linear $\Delta u \sim 10^{-5}$. The purple
and blue  curves correspond to system sizes $L_x=L_y=10^3$ and $L_x=L_y=10^2$ respectively,
and demonstrate that the plateau is a finite size effect. This regime appears to
be valid for approximately the interval $10^{-1.7} > u > 10^{-3.0}$.}
  \label{fig:fiMagOpScaling}
\end{figure}

The absolute values of the derivatives of $F_{M,\alpha}$ will be the same in the 
AFM and FM cases.
This can be seen by considering the fact that the functional dependence of
$F_{M,\alpha}$ on the driving parameter $u$ enters through the nearest neighbour
correlation functions, which are the same in both cases up to a negative sign.

The first and second derivatives of the QFI are given in
Fig.~\ref{fig:MagOpDouble}. We observe that the QFI susceptibility associated
with the magnetization operator exhibits a power law divergence when
approaching the critical point from the gapped side. When approaching the
critical point from the gapless side the transition appears first order. The
behaviour of the transition from the gapped side can be understood in light of the analysis in
Sec.~\ref{sec:QfiSuscepAndDivergingLengthScales}. 
Using the scaling hypothesis,
\begin{equation}
  \partial_u^2 F_{M,\alpha} (u) \sim \qty|u-u_c|^{\Delta_{M,\alpha}}
  \label{eq:ScalingHypothesisFiMagOp}
\end{equation}
we extract the following critical exponents for the second derivative of the QFI
for each spin component of the magnetization operator,
\begin{subequations}
  \begin{align}
    \Delta_{M,x} = \Delta_{M,y} &\approx -0.52(1) \\
    \Delta_{M,z} &\approx -0.50(1)
    \label{subeqs:fiMagOpScalingExponents}
  \end{align}
\end{subequations}
and can be seen in Fig.~\ref{fig:fiMagOpScaling} over a region from $10^{-1.7} >
u >10^{-3.0}$. At this point finite, finite size effects enter, and the scaling
ansatz is no longer valid. This leads to plateaus in the QFI susceptibility
which occur closer to the critical point for larger system sizes as seen from
the data in Fig.~(\ref{fig:fiMagOpScaling}) and
Fig.~(\ref{fig:bondScalingPlot}).

\begin{figure}[htp]
  \centering
  \includegraphics[width=0.5\textwidth]{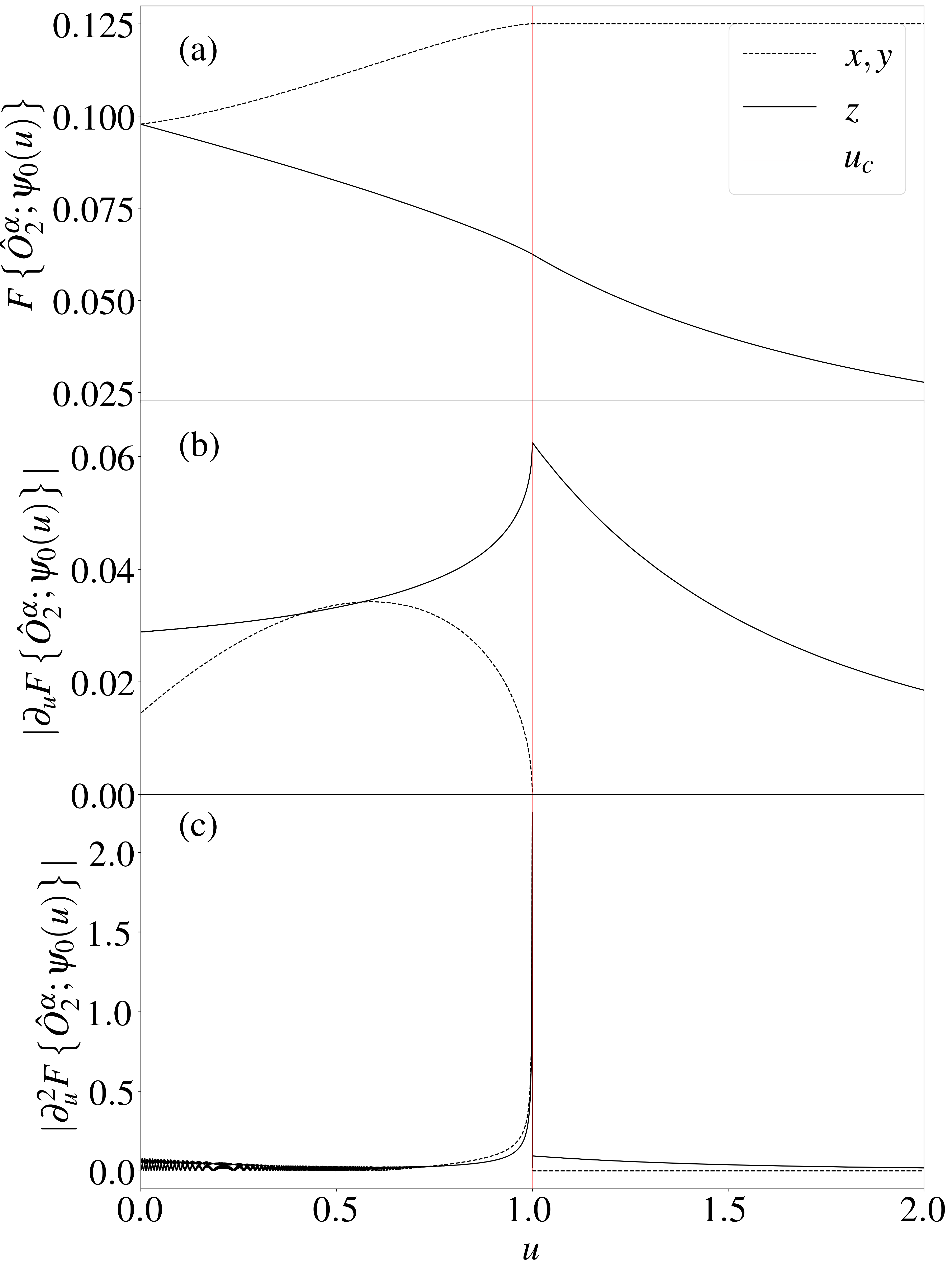}
  \caption{QFI associated with the bond correlation operator (a) with first (b)
    and second (c)
  derivatives. The critical point is marked with a solid vertical red line. The
second derivatives diverge, this time approaching the critical point from the
left (the gapless phase).}
  \label{fig:qfiBondOp}
\end{figure}
\subsection{Bond Correlation Operator}\label{sec:QfiSuscep:BondOp}

\begin{figure}[htp]
  \centering
  \includegraphics[width=0.5\textwidth]{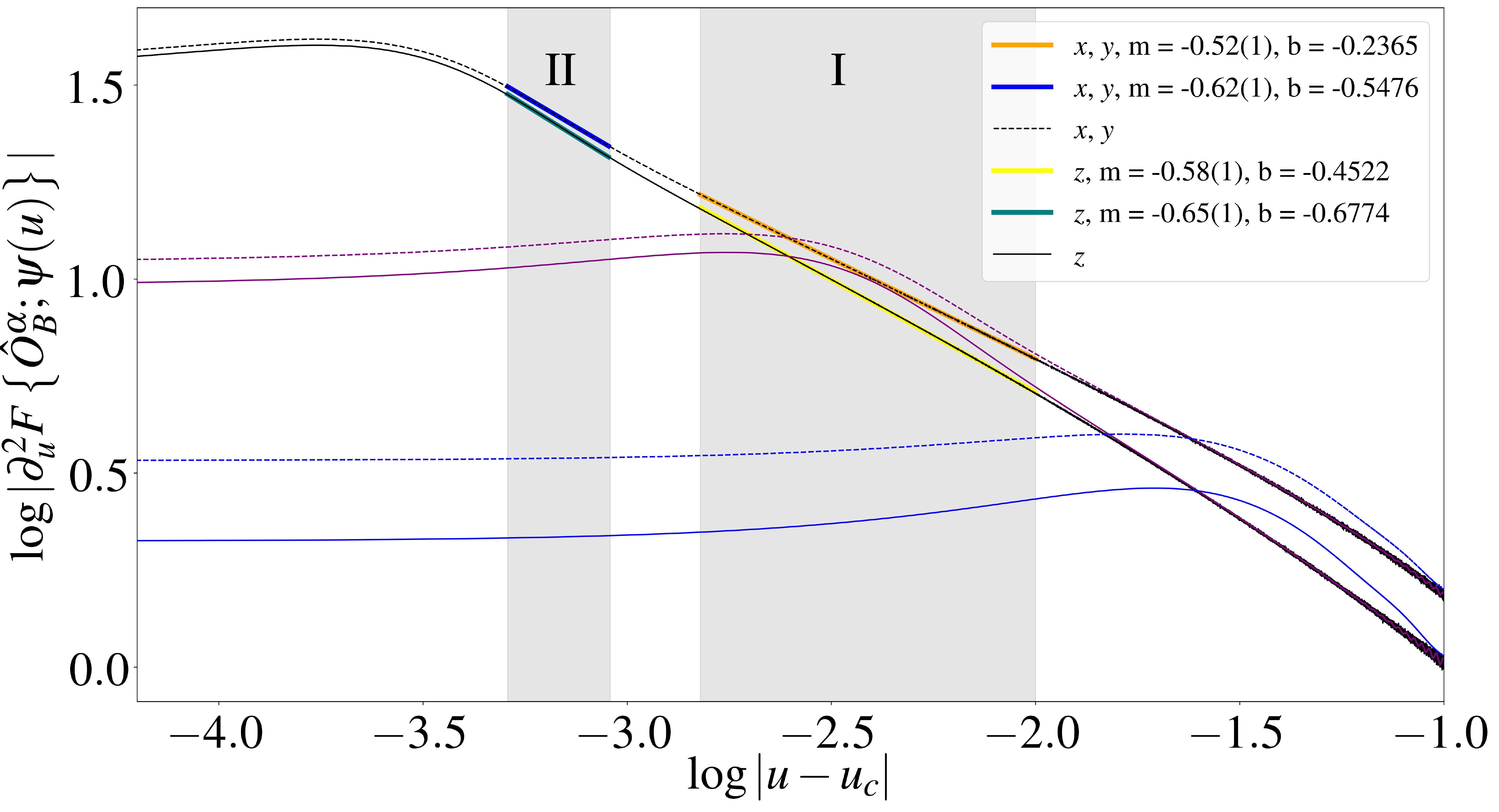}
  \caption{log-log plot of the second derivative of the QFI associated with the bond
  operator vs the distance from the critical point from the gapless phase
($u<1$). Results are the black curve are for $L_x=L_y=10^4$ with a line $\Delta
u\sim 10^{-5}$. The purple and blue curves correspond to square geometries of
size $L_x=L_y=10^3$ and $L_x=L_y=10^2$ respectively. The shaded regions (I) and (II) 
correspond to two regimes where we see linear scaling. The
first region is valid for approximately the interval, $10^{-1.9} > u >
10^{-2.8}$, while the second regime span approximately $10^{-3.0}> u>
10^{-3.3}$.}
  \label{fig:bondScalingPlot}
\end{figure}

We now turn to the QFI as parameterized by the bond correlation operator,
\begin{equation}
  \hat{O}_B^\alpha = \sum_r S_{r,A}^\alpha S_{r,B}^\alpha,
  \label{eq:BondOperatorDefn}
\end{equation}
with the corresponding QFI density given by, 
\begin{equation}
  F_{B,\alpha} = \frac{1}{2N}\sum_q \sin^2(2\theta_q) 
  \label{}
\end{equation}
In this case the AFM and FM cases are identical. 
We repeat the same analysis as for the magnetization operator as shown in
Fig.~\ref{fig:qfiBondOp} and ~\ref{fig:bondScalingPlot}. The QFI associated with the
bond operator along the $x$ and $y$ components converges to a constant value
immediately following the phase transition, while the QFI associated with the
$z$ component bond operator falls towards zero. This behaviour can be understood
by the fact that the Hamiltonian in the gapped phase is dominated by the Ising
exchange on the $z$ bonds. Consequently, the commutator between the Hamiltonian
and the bond operator approaches zero in the limit of $u\rightarrow\infty$. 

Assuming the same scaling ansatz as for the second derivative $F_{M,\alpha}$ we
find a crossover between two scaling regimes. The first regime is given by the critical
exponents, 
\begin{subequations}
  \begin{align}
    \Delta_{B,x}^{(\text{I})}=\Delta_{B,y}^{(\text{I})} &\approx -0.52(1) \\
    \Delta_{B,z}^{(\text{I})} &\approx -0.58(1) 
    \label{subeqs:BondOperatorsScalingI}
  \end{align}
\end{subequations}
which appears valid on the interval $10^{-1.9} > u >
10^{-2.8}$ and a second regime
characterized by the exponents,
\begin{subequations}
  \begin{align}
    \Delta_{B,x}^{(\text{II})}=\Delta_{B,y}^{(\text{II})} &\approx -0.62(1) \\
    \Delta_{B,z}^{(\text{II})} &\approx -0.65(1) 
    \label{subeqs:BondOperatorsScalingII}
  \end{align}
\end{subequations}
which appears to be valid on the interval $10^{-3.0}> u>
10^{-3.3}$. While the magnetization operator exhibits a divergence when approaching the
critical  point from the gapped phase, the bond operator exhibits a divergence
approaching the critical point from the gapless phase. 

The oscillatory behaviour on the gapless side of the transitions for both
QFI's is related to divergences in the QFI susceptibility due to points where
the denominator of the integrand goes to zero. In the gapped phase these points
are necessarily absent. 

\subsection{Diverging length scales}

Using the results of the previous section we can determine the scaling of the
divergence in the correlation length for $\qexp{c_{r_1,A}c_{r_2,B}}$ 
(using the divergence in the
magnetization operator), and for $\qexp{c_{r_1,A}c_{r_1,B}c_{r_2,A}c_{r_2,B}}$.

In light of Eq.~(\ref{eq:QfiSuscepDiverg}), we can now understand that the QFI
susceptibility associated with the magnetization operator diverges from the
gapped side due specifically to the divergence in that correlation function of
the matter sector Majorana's. On the gapless side of the transition, the
correlation function for the matter sector Majorana's is critical, and
consequently the second derivative of Eq.~(\ref{eq:CorrAnsatz}) is given
specifically by $\Xi(r,u)$ and contains no divergence.

Using
Eq.~(\ref{eq:QfiSuscepDiverg}), we may extract the scaling exponents for the
correlation length of the matter sector correlation functions in the $x$ and
$y$, and $z$ channels for $\partial_u^2F_{M,\alpha}$,
\begin{subequations}
  \begin{align}
    \nu_{M,x}=\nu_{M,y} &\approx 1.48(1) \\
    \nu_{M,z}&\approx 1.50(1)
    \label{}
  \end{align}
\end{subequations}
and for the two scaling regimes of $\partial_u^2F_{B,\alpha}$
for $u<u_c$. The first given by,
\begin{subequations}
  \begin{align}
    \nu_{B,x}^{\text{(I)}} = \nu_{B,y}^{\text{(I)}} &\approx 1.48(1) \\
    \nu_{B,z}^{\text{(I)}} &\approx 1.42(1),
    \label{}
  \end{align}
\end{subequations}
and the second by,
\begin{subequations}
  \begin{align}
    \nu_{B,x}^{\text{(II)}} = \nu_{B,y}^{\text{(II)}} &\approx 1.38(1) \\
    \nu_{B,z}^{\text{(II)}} &\approx 1.35(1),
    \label{}
  \end{align}
\end{subequations}

\section{Conclusion}\label{sec:Conclusion}

We have examined the QFI for the bond and magnetization operators in both the
gapped and gapless phases of the KHM and at the transition between these two
phases for fully ferromagnetic and fully anti-ferromagnetic couplings. The
second derivative of the QFI with respect to the magnetization operator is shown
to diverge when approaching the phase transition from the gapped side like at a
second order transition, while the QFI susceptibility approaching the critical
point from the gapless side appears first order. Conversely we find that the QFI 
susceptibility associated with the bond operator
diverges like a second order transition when approaching the critical point from
the gapless side, and like a first order transition when approaching the
critical point from the gapped side. 

In both cases, the divergences in the QFI susceptibility can be associated to
diverging length scales in the two point correlators of the local generators of
the QFI. For the particular case of the KHM, these divergences can be linked to
diverging length scales in the matter sector Majorana's, even when the physical
spin-spin correlation functions are truncated (as in the case of the two point
correlation function). The implication is that presence of the topological phase
transition between the gapped and gapless phases may be detected experimentally
at low temperatures.  

The has been related work examining the Geometric phase associated with a twist operator
acting on both sites.~\cite{lian2012geometric} We note that the critical
exponents presented in Eq.~\ref{subeqs:BondOperatorsScalingII} for scaling
regime (II) of the bond operators are within the margin of error of those in
Ref.~\cite{lian2012geometric}. Geometrically, the QFI that we compute with
respect to the bond operator is the diagonal component of the quantum geometric
tensor~\cite{provost1980riemannian}. The imaginary component of this tensor
corresponds to the Berry curvature, while the real component corresponds to the
notion of distance induced by the distinguishability of states. In
Ref.~\cite{PhysRevLett.121.170401}, the connection between these two components
of metric is discussed. The implication is that the geometry detected by the
Berry phase is intimately related to the geometry of distinguishability, opening
the prospect of experimentally measuring the Berry phase in condensed matter
systems. Extracting the full quantum metric tensor has recently been achieved in
cold atom systems.~\cite{PhysRevLett.122.210401}

The QFI associated with the magnetization operator in the fully anti-ferromagnetic
phase is shown to be greater than for the fully ferromagnetic phase, as one
would expect from the tendency of the anti-ferromagnetic coupling to produce
spin singlets on the bonds. In the gapped phase defined by large $K^\gamma$
coupling, the QFI associated with the
bond operator converges to a constant value for the QFI generated by the
transverse spin components (specifically the x and y components in our
analysis).

It is shown in ~\cite{liu2014fidelity} that the QFI is proportional to the
FS if the operator parameterizing the QFI is the same as the
operator that generates the change in parameter for the ground state. 
This implies that the QFI for the bond operator from the gapless side of the
transition is proportional the FS calculated
in Ref.~\cite{yang2008fidelity}, however we do not find this to be the case.
This may be due to the fact that the ground state used in
Ref.~\cite{yang2008fidelity} differs from that used by Kitaev in
Ref.~\cite{AnyonsExactlySolved}, which is the one we employ here. 
Understanding the details  of the connection between the QFI and the fidelity
susceptibility warrants further investigation. We also note that the definition of the QFI may not be unique when a
Hamiltonian posseses a degenerate ground state manifold. In particular one can
imagine a situation where the generator of the QFI lifts the degeneracy of this
manifold, affecting the results. While in our calculation this ambiguity is not present
in the gapped phase, it may affect the results in the gapless phase. 

Future research is warranted to examine the behaviour of the QFI at finite
temperatures around the critical point, where the ground state scaling will be
modified by finite temperature effects. The connection between the finite
temperature scaling and the length scale of the Majorana fermions in this case may
offer insight into the details of candidate Kitaev spin liquid phases in
materials where the pure Kitaev Hamiltonian is modifed by material relevant
terms.~\cite{rau2014generic}

\onecolumngrid
\appendix

\section{Scaling behaviour of QFI Susceptibility}

 Let's work specifically on the case of pure states and unitary QFI. 
The generator of the QFI is most generally given by,
\begin{equation}
  \hat{O} = \sum_r\hat{O}_r
  \label{}
\end{equation}
where $\hat{O}_r$ is an operator associated with the site located at $r$. We
assume that $r$ is contiguous and local, that is it encompasses a finite number of degrees of
freedom all lying within a distance $\ell$ from the site $r$. We consider a
state $\psi$  that depends on some parameter $u$ that drives a phase transition
at a value $u_c = 1$. 
\begin{equation}
  f\{\hat{O},\psi(u)\} =\frac{1}{N} \sum_{r_1,r_2} \langle\hat{O}_{r_1}\hat{O}_{r_2}\rangle_\psi -
  \langle\hat{O}_{r_1}\rangle_\psi\langle\hat{O}_{r_2}\rangle_\psi
  = \frac{1}{N}\sum_{r_1,r_2}C_{r_1,r_2}(u)
  \label{}
\end{equation}
Let's assume that the model is translation invariant and define $r\coloneqq
\qty|r_1-r_2|$. In general, we may assume that the connected correlation functions can be
fit to the following form,
\begin{equation}
  C_r(u) = \Xi(r,u)r^{-a}e^{-\frac{r}{\xi(u)}}
\end{equation}
where $a$ depends on the phase (i.e. is assumed independent of the driving
parameter), and $\xi$ is the correlation length, taken to be a function of the
parameter $u$ (we hereafter drop the explicit dependence). The function
$\Xi(r,u)$ is assumed to be a smooth function of $r$ and of the parameter $u$
within a particular phase (though not necessarily smooth at the phase boundary). 
The divergence in the
second derivative of the QFI must emerge from a divergence in the two point
correlation functions. We therefore consider the second derivative of
Eq.~(\ref{eq:CorrAnsatz})
\begin{align}
  \partial_u^2C_r(u) &= 
  \partial_u
  \left(
    \Xi(r,u)r^{1-a}\xi^{-2}\partial_u\xi e^{-\frac{r}{\xi}} 
    + \partial_u\Xi(r,u)r^{-a}e^{-\frac{r}{\xi}}
  \right) \nonumber \\
  &= 
  -2r^{1-a}\Xi(r,u)\xi^{-3}\left(\partial_u\xi\right)^2 e^{-\frac{r}{\xi}}
  + r^{1-a}\Xi(r,u)\xi^{-2}\partial_u^2\xi e^{-\frac{r}{\xi}} 
  + r^{2-a}\Xi(r,u)\xi^{-4}\left(\partial_u\xi\right)^2
  e^{-\frac{r}{\xi}}\nonumber \\ 
  &+\partial_u^2\Xi(r,u)r^{-a}e^{-\frac{r}{\xi}}
  + \partial_u\Xi(r,u)r^{1-a}\xi^{-2}\partial_u\xi e^{-\frac{r}{\xi}}
  \label{eq:CorrSecDeriv}
\end{align}
Naively, the correlation length is expected to diverge at the critical point.
let $\tilde{u}=\qty|u-u_c|$ be the distance from the critical point. Then the
correlation length goes as,
\begin{equation}
  \xi \sim \tilde{u}^{-\nu}
  \label{}
\end{equation}
This ansatz may be used to infer the scaling relations for the derivatives of
the correlation length,
\begin{subequations}
\begin{align}
  \partial_u\xi &\sim -\nu\tilde{u}^{-(\nu+1)} \\
  \partial_u^2\xi &\sim \nu(\nu+1) \tilde{u}^{-(\nu+2)}
  \label{}
\end{align}
\end{subequations}
Substituting this into Eq.~(\ref{eq:CorrSecDeriv}), gives,
\begin{align}
  \partial_u^2C_r(u) &= 
  - 2\Xi(r,u)r^{1-a}\tilde{u}^{3\nu}\nu^2\tilde{u}^{-2(\nu+1)}e^{-\frac{r}{\xi}}
  + \Xi(r,u)r^{1-a} \tilde{u}^{2\nu}v(v+1)\tilde{u}^{-(\nu+1)} e^{-\frac{r}{\xi}} 
  + \Xi(r,u)r^{2-a}\tilde{u}^{4\nu}\nu^2\tilde{u}^{-2(\nu+1)}e^{-\frac{r}{\xi}} \nonumber \\
  &+\partial_u^2\Xi(r,u)r^{-a}e^{-\frac{r}{\xi}}
  + \partial_u\Xi(r,u)r^{1-a}\tilde{u}^{2\nu}(-\nu)\tilde{u}^{-(\nu+1)}
  e^{-\frac{r}{\xi}} \nonumber \\
    &= e^{-\frac{r}{\xi}}
    \left(
    -2\Xi(r,u)r^{1-a} \tilde{u}^{\nu - 2}\nu^2 
    + \Xi(r,u)r^{1-a} \tilde{u}^{\nu - 2}\nu(\nu+1)
    + \Xi(r,u)r^{2-a} \tilde{u}^{2\nu-2} \right.\nonumber \\
    &+\left.\partial_u^2\Xi(r,u)r^{-a} 
    -\nu\partial_u\Xi(r,u)r^{1-a}\tilde{u}^{\nu-2}
    \right)
  \label{}
\end{align}
We can now pull out the divergence associated with the proximity to the critical
point. 
\begin{align}
  \partial_u^2 C_r(u) &= \tilde{u}^{\nu-2}e^{-\frac{r}{\xi}}
  \left(
    -2\Xi(r,u)r^{1-a} \nu^2 
    + \Xi(r,u)r^{1-a} \nu(\nu+1)
    + \Xi(r,u)r^{2-a} \tilde{u}^{\nu} \right.\nonumber \\
    &+\left.\tilde{u}^{2-\nu}\partial_u^2\Xi(r,u)r^{-a} 
    -\nu\partial_u\Xi(r,u)r^{1-a}
  \right)
\end{align}
The scaling behaviour of the QFI susceptibility is thus given by,
\begin{equation}
  \partial_u^2f\{\hat{O},\psi\}=\tilde{u}^{\nu-2}\zeta(r,u)
  \label{}
\end{equation}
We define $\Delta_{\hat{O}} = \nu- 2$ as the scaling of the QFI with proximity
to the critical point. The y-intercept on the log-log plot will be given by the
non-universal function $\zeta(r,u)$.

\section{Calculating Variances}\label{app:CalcVariances}

\subsection{Solution of the Kitaev Model}\label{app:CalcVariance:SolnKitaev}

We adopt the approach of~\cite{baskaran2007exact,knolle2015dynamics}, where the Majorana degrees of
freedom are recombined into Dirac fermions, with three \emph{bond} fermions,
\begin{subequations}
  \begin{align}
    b_{r,A}^\gamma &= \frac{1}{2}\left(\beta_r^\gamma +
    (\beta_r^\gamma)^\dagger\right) \\
    b_{r,B}^\gamma &= \frac{1}{2i}\left(\beta_r^\gamma -
    (\beta_r^\gamma)^\dagger\right) \\
  \end{align}
\end{subequations}\label{eq:DefnBondFerm}
and one \emph{matter} fermion
\begin{subequations}
\begin{align}
  c_{r,A} &= \frac{1}{2}\left(f_r + f_r^\dagger\right) \\
    c_{r,B} &= \frac{1}{2}\left(f_r - f_r^\dagger\right) 
  \end{align}
\end{subequations}\label{eq:DefnMatterFerm}
The bond fermions are not present in the Hamiltonian, since we simply replace
the bond operators with the eigenvalues of the standard gauge configuration
($u_{j,k}=1$). The resulting Hamiltonian is quadratic in the matter fermions and
translation invariant. It can be diagonalized first by mapping each matter
fermion to momentum space, $f_r = \frac{1}{\sqrt{N}}\sum_q e^{iq\cdot r} f_q$,
and then applying the Bogoliubov rotation, $f_q = \cos(\theta_q)a_q +
i\sin(\theta_q)a_{-q}^\dagger$, where $\theta_q$ is defined by,
\begin{equation}
  \tan(2\theta_q) = \frac{K^x\cos(q_x) + K^y\cos(q_y) + K^z}{K^x\sin(q_x) +
  K^y\sin(q_y)}.
  \label{}
\end{equation}

\newcommand{\mago}{\hat{O}^\alpha_\text{Mag}}
\newcommand{\twos}{\hat{O}^\alpha_\text{2-Site}}
\subsection{Magnetization Operator}
Begin with, 
\begin{equation}
  \mago = \sum_{r} \spin{\alpha}_r 
  \label{}
\end{equation}
The variance is given generally by, 
\begin{equation}
  \qvar{\hat{O}} = \qexp{O^2} - \qexp{O}^2,
  \label{}
\end{equation}
which, for the magnetization operator gives,
\begin{equation}
  \qvar{\mago} = \sum_{r_1,r_2}
  \qexp{\spin{\alpha}_{r_1}\spin{\alpha}_{r_2}}
  \label{}
\end{equation}
Using translation invariance and converting the Majorana representation, this
expression can be given as,
\begin{align}
  \qvar{\hat{O}^\alpha_{\text{Mag}}} &= N \sum_{r}
  \qexp{\spin{\alpha}_0\spin{\alpha}_r}
  \label{}
\end{align}
In the Kitaev model, the two  point correlator is zero for all values of $r$
except nearest neighbours. Thus the sum above can be reduced to, 
\begin{equation}
  \qvar{\hat{O}^\alpha_{\text{Mag}}} = N \left( \frac{1}{4} +
  \qexp{\spin{\alpha}_{0,A}\spin{\alpha}_{0,B}}\right)
  \label{}
\end{equation}
Thus we only need to calculate the nearest neighbour correlation function, 
\begin{align}
  \qexp{\spin{\alpha}_{0,A}\spin{\alpha}_{0,B}}
  &= \frac{1}{4} \qexp{\sigma^\alpha_{0,A}\sigma^\alpha_{0,B}}\nonumber \\
  &= \frac{1}{4} \qexp{(ib_{0,A}^\alpha c_{0,A})(ib_{0,B}^\alpha c_{0,B})}
  \nonumber \\
  &= \frac{1}{4}
  \bra{\mathcal{F}}b_{0,A}^\alpha b_{0,B}^\alpha\ket{\mathcal{F}}
  \bra{\mathcal{M}}c_{0,A}c_{0,B}\ket{\mathcal{M}}\nonumber \\
  &= \frac{1}{4}\bra{\mathcal{F}}(-i)(2\hat{n}^{\beta^\alpha}_0 - 1)\ket{\mathcal{F}}
  \bra{\mathcal{M}}(-i) (2\hat{n}^f_0 - 1)\ket{\mathcal{F}} \nonumber \\
  &= -\frac{1}{4}\left(2\qexp{\hat{n}^f_0} - 1\right) \nonumber \\
  &= -\frac{1}{4}\left(2\frac{1}{N} \sum_{q_1,q_2} \qexp{f^\dagger_{q_1}f_{q_2}} -
  1\right)\nonumber \\
  &= -\frac{1}{4}\left(\frac{2}{N} \sum_{q} \sin^2(\theta_q) - 1\right)
  \label{}
\end{align}
The QFI density is four times the variance divided by the system size. Thus,
\begin{equation}
  f\left\{\mago\right\} = 1 +
  \qexp{\sigma_{0,A}^\alpha\sigma_{0,B}^\alpha}
  \label{}
\end{equation}

\subsection{Bond Operator}
\begin{align}
  \text{Var}(\twos) &= \qexp{(\twos)^2} - \qexp{\twos}^2 \nonumber \\
  &= \sum_{r_1,r_2} \qexp{\spin{\alpha}_{r_1 A}\spin{\alpha}_{r_1 B}
                          \spin{\alpha}_{r_2 A}\spin{\alpha}_{r_2 A}}
    - \left(\sum_r \spin{\alpha}_{r_1 A}\spin{\alpha}_{r_1 B}\right)^2 \nonumber
    \\
    &= \frac{1}{16}\left(\sum_{r_1,r_2} 
    \qexp{b_{r_1 A}^\alpha b_{r_1 B}^\alpha b_{r_2 A}^\alpha b_{r_2 B}^\alpha}
    \qexp{c_{r_1 A}c_{r_1 B}c_{r_2 A}c_{r_2 B}} - \left(\sum_r
    -\qexp{b_{r A}^\alpha b_{r B}^\alpha}\qexp{c_{r A}c_{r B}}\right)^2
    \right)
  \label{}
\end{align}
We can compute the flux sector expectation values easily, 
\begin{align}
  \qexp{b_{r_1 A}^\alpha b_{r_1 B}^\alpha b_{r_2 A}^\alpha b_{r_2 B}^\alpha}
  &= (-i)^2\qexp{(\beta_{r_1} + \beta_{r_1}^\dagger)(\beta_{r_1}-\beta_{r_1}^\dagger)
           (\beta_{r_2} + \beta_{r_2}^\dagger)(\beta_{r_2}-\beta_{r_2}^\dagger)}
           \nonumber \\ 
  &= -\qexp{(2n_{r_1} - 1)(2n_{r_2} - 1)} \nonumber \\
  &= -1
  \label{}
\end{align}
Where the last line
follows from the fact that the ground state in the standard flux configuration
is defined by $u_r = 2n_r - 1 = 1$. Similarly we find,
\begin{align}
  \qexp{b_{rA}^\alpha b_{rB}^\alpha} &= (-i)\qexp{(2n_r -1)} \nonumber \\
                                     &= (-i)
  \label{}
\end{align}
The variance is therefore,
\begin{align}
  \text{Var}(\twos) &= \frac{1}{16}\left(\sum_{r_1,r_2}
  (-1)\qexp{c_{r_1A}c_{r_1B}c_{r_2A}c_{r_2B}}
  -\left( \sum_r(i)\qexp{c_{rA}c_{rB}}\right)^2  
  \right)
  \nonumber \\ 
  &= \frac{1}{16}\left(
  -\sum_{r_1,r_2}
  \qexp{c_{r_1 A}c_{r_1 B}c_{r_2 A}c_{r_2 B}}
  + 
  \sum_{r_1,r_2} \qexp{c_{r_1 A} c_{r_1 B}}\qexp{c_{r_2 A} c_{r_2 B}} 
  \right) \nonumber \\
  &= \frac{1}{16}\left(
   \sum_{r_1,r_2} 
   -\qexp{c_{r_1 A}c_{r_1 B}}\qexp{c_{r_2 A}c_{r_2 B}} 
   +\qexp{c_{r_1 A}c_{r_2 A}}\qexp{c_{r_1 B}c_{r_2 B}}
   -\qexp{c_{r_1 A}c_{r_2 B}}\qexp{c_{r_1 B}c_{r_2 A}} \right. \nonumber \\
    &+\left.\qexp{c_{r_1 A}c_{r_1 B}}\qexp{c_{r_2 A}c_{r_2 B}} 
 \right) \nonumber \\
  &= \frac{1}{16}
  \left(
    \sum_{r_1,r_2} 
    \qexp{c_{r_1 A}c_{r_2 A}}\qexp{c_{r_1 B}c_{r_2 B}}
   -\qexp{c_{r_1 A}c_{r_2 B}}\qexp{c_{r_1 B}c_{r_2 A}} 
  \right)
  \label{}
\end{align}
We now need only evaluate the two point correlators above. For the first term we
have,
\begin{align}
  \qexp{c_{r_1 A} c_{r_2 A}} &= 
  \frac{1}{N}\sum_{q_1,q_2} e^{iq_1r_1} e^{iq_2r_2}\qexp{(f_{q_1} + f_{q_1}^\dagger)}
                 \qexp{(f_{q_2} + f_{q_2}^\dagger)} \nonumber \\
                 &= \frac{1}{N}\sum_{q_1,q_2}e^{iq_1r_1} e^{iq_2r_2} 
            \qexp{f_{q_1}f_{q_2}+f_{q_1}f_{q_2}^\dagger + f_{q_1}^\dagger
            f_{q_2} + f_{q_1}^\dagger f_{q_2}^\dagger} \nonumber \\
            &=\frac{1}{N}\sum_{q_1,q_2}e^{iq_1r_1} e^{iq_2r_2}
            \qexp{f_{q_1}f_{q_2} + f_{q_1}^\dagger f_{q_2}^\dagger +
            \delta_{q_1,q_2}}
  \label{}
\end{align}
We can see that,
\begin{subequations}
\begin{align}
  \qexp{f_{q_1}f_{q_2}} &= i\cos(\theta_{q_1})\sin(\theta_{q_2})\qexp{a_{q_1}
  a_{-q_2}^\dagger} \nonumber \\
  &= i\cos(\theta_{q_1})\sin(\theta_{q_2}) \delta_{q_1,-q_2} \nonumber \\
  \qexp{f_{q_1}^\dagger f_{q_2}^\dagger} &=
  (-i)\cos(\theta_{q_1})\sin(\theta_{q_2}) \qexp{a_{-q_1}a_{q_2}^\dagger}
  \nonumber \\
  &= (-i) \cos(\theta_{q_1})\sin(\theta_{q_2})\delta_{-q_1,q_2}
  \label{}
\end{align}
\end{subequations}
Consequently, 
\begin{align}
  \qexp{c_{r_1 A} c_{r_2 A}} &= 
  \frac{1}{N} \sum_{q_1,q_2} e^{iq_1r_1}e^{iq_2r_2} \delta_{q_1,q_2} \nonumber
  \\
  &= \frac{1}{N} \sum_q e^{iq(r_1-r_2)} \nonumber \\
  &= \delta_{r_1,r_2}
  \label{}
\end{align}
Similarly we may show that, 
\begin{equation} 
  \qexp{c_{r_1 B} c_{r_2 B}} = \delta_{r_1,r_2} 
  \label{}
\end{equation}
For the second term we begin with, 
\begin{align}
  \qexp{c_{r_1A}c_{r_2B}}&=
  \frac{1}{N}\sum_{q_1,q_2}e^{iq_1r_1}e^{iq_2r_2}
  (-i)\qexp{(f_{q_1}+f_{q_1}^\dagger)(f_{q_2}-f_{q_2}^\dagger} \nonumber \\
  &= \frac{1}{N}\sum_{q_1,q_2}e^{iq_1r_1}e^{iq_2r_2}(-i)
  \qexp{f_{q_1}f_{q_2}-f_{q_1}f_{q_2}^\dagger + f_{q_1}^\dagger f_{q_2}
  -f_{q_1}^\dagger f_{q_2}^\dagger } \nonumber \\
  &= \frac{(-i)}{N} \sum_{q_1,q_2} e^{iq_1r_1}e^{iq_2r_2} 
     \left(\qexp{f_{q_1}f_{q_2} - f_{q_1}^\dagger f_{q_2}^\dagger 
                +f_{q_1}^\dagger f_{q_2} - f_{q_1}f_{q_2}^\dagger}\right)
                \nonumber \\
  &= \frac{(-i)}{N} \sum_{q_1,q_2} e^{iq_1r_1}e^{iq_2r_2}
  \left(\qexp{f_{q_1}f_{q_2}-f_{q_1}^\dagger f_{q_2}^\dagger} 
       +\qexp{2f_{q_1}^\dagger f_{q_2} - 1}\right) \nonumber \\
       &= \frac{(-i)}{N}\sum_{q_1,q_2}e^{iq_1r_1}e^{iq_2r_2}
       \left(i\cos(\theta_{q_1})\sin(\theta_{q_2})(\delta_{q_1,-q_2} +
       \delta_{-q_1,q_2}) +
       2\sin(\theta_{q_1})\sin(\theta_{q_2})\delta_{q_1,q_2} - 1\right)\nonumber
       \\
       &=\frac{(-i)}{N}\sum_{q} 2ie^{-iq(r_1-r_2)}\cos(\theta_q)\sin(\theta_{q}) 
       +\frac{(-i)}{N}\sum_{q} (2\sin^2(\theta_q) - 1)\nonumber \\
       &= \frac{1}{N}\sum_q e^{-iq(r_1-r_2)}2\cos(\theta_q)\sin(\theta_q)
       +\frac{i}{N}\sum_qe^{iq(r_1+r_2)} \cos(2\theta_q) \nonumber \\
       &= \frac{1}{N}\sum_q e^{-iq(r_1-r_2)} \sin(2\theta_q) 
          + \frac{i}{N}\sum_qe^{iq(r_1+r_2)} \cos(2\theta_q)
  \label{}
\end{align}
The final two point correlator is given by,
\begin{align}
  \qexp{c_{r_1B}c_{r_2A}} &= \frac{1}{N}\sum_{q_1,q_2}e^{iq_1r_1}e^{iq_2r_2} 
  (-i)\qexp{(f_{q_1}-f_{q_1}^\dagger)(f_{q_2}+f_{q_2}^\dagger)} \nonumber \\
  &=\frac{(-i)}{N}\sum_{q_1,q_2}e^{iq_1r_1}e^{iq_2r_2}
  \left(
    \qexp{f_{q_1}f_{q_2} - f_{q_1}^\dagger f_{q_2}^\dagger}
    -\qexp{2f_{q_1}^\dagger f_{q_2} - 1} 
  \right)\nonumber \\
  &=\frac{(-i)}{N}\sum_{q_1,q_2}e^{iq_1r_1}e^{iq_2r_2}
  \left(i\cos(\theta_{q_1})\sin(\theta_{q_2})(\delta_{q_1,-q_2} +
  \delta_{-q_1,q_2}) - (2\sin(\theta_{q_1})\sin(\theta_{q_2})\delta_{q_1,q_2} -
  1)
  \right) \nonumber \\
  &= \frac{(-i)}{N}\sum_q 2ie^{-iq(r_1-r_2)}\cos(\theta_q)\sin(\theta_q)
  -\frac{(-i)}{N}\sum_q e^{iq(r_1+r_2)}(2\sin^2(\theta_q) - 1) \nonumber \\
  &= \frac{1}{N}\sum_qe^{-iq(r_1-r_2)}\sin(2\theta_q) -
  \frac{i}{N}\sum_qe^{iq(r_1+r_2)}\cos(2\theta_q)
  \label{}
\end{align}
The final term is therefore a product of differences,
\begin{align}
  \qexp{c_{r_1B}c_{r_2A}}\qexp{c_{r_1A}c_{r_2B}}
  &=
  \frac{1}{N^2}\sum_{q_1,q_2}\left(e^{-i(q_1+q_2)(r_1-r_2)}
  \sin(2\theta_{q_1})\sin(2\theta_{q_2})
  +e^{i(q_1+q_2)(r_1+r_2)}\cos(2\theta_{q_1})\cos(2\theta_{q_2})\right) \nonumber \\
  &= \frac{1}{N^2}\sum_{q_1,q_2} \left(e^{ir_1(q_1+q_2)}
  e^{-ir_2(q_1+q_2)}\sin(2\theta_{q_1})\sin(2\theta_{q_2})
  + e^{ir_1(q_1+q_2)}e^{ir_2(q_1+q_2)}\cos(2\theta_{q_1})\cos(2\theta_{q_2})
 \right)
  \label{}
\end{align}
Under the summation, we may extract the delta functions,
\begin{align}
  \sum_{r_1,r_2}\qexp{c_{r_1B}c_{r_2A}}\qexp{c_{r_1A}c_{r_2B}} 
  &= \frac{1}{N}\sum_{r_2}\sum_{q_1,q_2} \delta_{q_1,-q_2}e^{-ir_2(q_1+q_2)}
  \sin(2\theta_{q_1})\sin(2\theta_{q_2}) + \delta_{q_1,-q_2}
  e^{ir_2(q_1+q_2)}\cos(2\theta_{q_1})\cos(2\theta_{q_2})  \nonumber \\
&= \sum_q \cos^2(2\theta_q) - \sin^2(2\theta_q)
  \label{}
\end{align}
We are now ready to return to our original expression for the variance  which
reads,
\begin{align}
  \text{Var}(\twos) &= \frac{1}{16}\left(N - \sum_q (\cos^2(2\theta_q) -
  \sin^2(2\theta_q))\right)  \nonumber \\
  &= \frac{1}{16}\sum_q\left(1-\cos^2(2\theta_q) +
  \sin^2(2\theta_q)\right) \nonumber \\
  &= \frac{1}{8} \sum_q\sin^2(2\theta_q)
  \label{}
\end{align}
Thus the associated QFI density is,
\begin{equation}
  \mathcal{F}(\twos) = 4\frac{\text{Var}(\twos)}{N} = 
  \frac{1}{2}\frac{1}{N}\sum_q\sin^2(\theta_q)  \label{}
\end{equation}
passing to the continuum limit we find, 
\begin{equation}
  \mathcal{F}(\twos) = \frac{1}{2}\int_{\text{BZ}}\sin^2(2\theta_q) \dd^2q 
  \label{}
\end{equation}

\section{Derivatives of the QFI Susceptibility Momentum Density}

In order to gain more insight into the divergence in $F_{M,\alpha}$, we define
the QFI \emph{momentum density} for the magnetization operator, $f_{M,\alpha}$
via,
\begin{equation}
  F_{M,\alpha} = \sum_q\left(\frac{1}{N} - \cos(2\theta_q)\right) = \sum_q
  f_{M,\alpha}(q;u)
  \label{}
\end{equation}
We may explicitly evaluate the first and second derivative of this quantity for
each spin component, giving, 
\begin{subequations}
  \begin{align}
    \partial_uf_{M,x} &= 
    \frac{
      \Delta_x^2\cos(q_y) - \varepsilon_x\Delta_x\sin(q_y)
    }{
      \left(\varepsilon_x^2 + \Delta_x^2\right)^\frac{3}{2}
    }\\
    \partial_uf_{M,y} &= 
    \frac{
      \Delta_y^2\cos(q_x) - \varepsilon_y\Delta_y\sin(q_x)
    }{
      \left(\varepsilon_y^2 + \Delta_y^2\right)^\frac{3}{2}
    }\\
    \partial_uf_{M,z} &=
    \frac{\Delta_z^2}{\left(\varepsilon_z^2 + \Delta_z^2\right)^\frac{3}{2}}
    \label{}
  \end{align}
\end{subequations}
for the first derivatives, and
\begin{subequations}
  \begin{align}
  \partial_u^2 f_{M,x} &=
  \frac{
    -3\varepsilon_x\Delta_x^2\cos^2(q_y)
    +\Delta_x(3(\varepsilon_x^2-\Delta_x^2) +
    \varepsilon_x+\Delta_x)\sin(q_y)\cos(q_y)
    +\varepsilon_x(2\Delta_x^2-\varepsilon_x^2)\sin^2(q_y)
  }{
    \left(\varepsilon_x^2+\Delta_x^2\right)^\frac{5}{2}
  } \\
  \partial_u^2 f_{M,y} &=
  \frac{
    -3\varepsilon_y\Delta_y^2\cos^2(q_x)
    +\Delta_y(3(\varepsilon_y^2-\Delta_y^2) +
    \varepsilon_y+\Delta_y)\sin(q_x)\cos(q_x)
    +\varepsilon_y(2\Delta_y^2-\varepsilon_y^2)\sin^2(q_x)
  }{
    \left(\varepsilon_y^2+\Delta_y^2\right)^\frac{5}{2}
  } \\
  \partial_u^2 f_{M,z} &=
  \frac{
    -3\varepsilon_z\Delta_z^2
  }{
    \left(\varepsilon_z^2+\Delta_z^2\right)^\frac{5}{2}
  }
    \label{}
  \end{align}
\end{subequations}
for the second derivatives.
We again compute explicitly the first and second derivatives of the QFI with
respect to the driving parameter by rewriting the QFI in terms of an integral
over a QFI density,
\begin{equation}
  F_{B,\alpha} = \int_{\text{BZ}} f_{B,\alpha}(q;u)\nonumber
  \label{}
\end{equation}
finding,
\begin{subequations}
\begin{align}
  \partial_u f_{B,x} &= 
  \frac{
    \Delta_x\varepsilon_x(\varepsilon_x\sin(q_y) -
    \Delta_x\cos(q_y))
  }{
    (\varepsilon_x^2+\Delta_x^2)^2
  }\\
  \partial_u f_{B,y} &= 
  \frac{
    \Delta_y\varepsilon_y(\varepsilon_y\sin(q_x) -
    \Delta_y\cos(q_x))
  }{
    (\varepsilon_y^2+\Delta_y^2)^2
  }\\
  \partial_u f_{B,z} &= 
  \frac{
    \Delta_z^2\varepsilon_z
  }{
    (\varepsilon_z^2+\Delta_z^2)^2
  }
  \label{}
\end{align}
\end{subequations}
for the first derivatives and
\begin{subequations}
\begin{align}
  \partial_u^2 f_{B,x} &=
  \frac{
    \left(\varepsilon_x\sin(q_y)-\Delta_x\cos(q_y)\right)
    \left(
      \left(
        \Delta_x^2-3\varepsilon_x^2
      \right)\Delta_x\cos(q_y)
      +
      \left(
        \varepsilon_x^2-3\Delta_x^2
      \right)\varepsilon_x\sin(q_y)
    \right)
  }{
    \left(\varepsilon_x^2+\Delta_x^2\right)^3
  } \\
  \partial_u^2 f_{B,y} &=
  \frac{
    \left(\varepsilon_y\sin(q_x)-\Delta_y\cos(q_x)\right)
    \left(
      \left(
        \Delta_y^2-3\varepsilon_y^2
      \right)\Delta_y\cos(q_x)
      +
      \left(
        \varepsilon_y^2-3\Delta_y^2
      \right)\varepsilon_y\sin(q_x)
    \right)
  }{
    \left(\varepsilon_y^2+\Delta_y^2\right)^3
  } \\
  \partial_u^2 f_{B,z} &=
  \frac{
    \Delta_z^2(\Delta_z^2-3\varepsilon_z^2)
  }{\left(\varepsilon_z^2+\Delta_z^2\right)^3}
  \label{}
\end{align}
\end{subequations}

\twocolumngrid\bibliography{ref}

\end{document}